\documentclass{egpubl}
\usepackage{eg2024}
\pdfoutput=1
\ConferencePaper        % uncomment for (final) Conference Paper
\CGFStandardLicense
\usepackage[T1]{fontenc}
\usepackage{dfadobe}  

\usepackage{cite}  % comment out for biblatex with backend=biber
\BibtexOrBiblatex
\electronicVersion
\PrintedOrElectronic
\ifpdf \usepackage[pdftex]{graphicx} \pdfcompresslevel=9
\else \usepackage[dvips]{graphicx} \fi

\usepackage{egweblnk} 
\title[Generate Coherent Rays Directly]%
{Generate Coherent Rays Directly}

\author[]
{\parbox{\textwidth}{\centering Fengqi Liu$^{1}$, 
		Zaonan Tan$^{1}$, Weilai Xiang$^{1}$, Chenhao Lu$^{1}$, Dan Li$^{1}$\thanks{Corresponding author},
		Xu Gong$^{2}$, Yulong Shi$^{2}$, Songnan Shi$^{2}$, Qilong Kou$^{2}$, Bo Hu$^{2}$        
        }
        \\
{\parbox{\textwidth}{\centering $^1$Huazhong University of Science and Technology, China\\
         $^2$Tencent
       }
}
}
\begin{document}

%\teaser{
% \includegraphics[width=0.9\linewidth]{eg_new}
% \centering
%  \caption{New EG Logo}
%\label{fig:teaser}
%}

\maketitle
%-------------------------------------------------------------------------
\begin{abstract}
The path tracing method generates incoherent rays by randomly sampling directions. This randomness makes it unsuitable for modern processor architectures that rely on coherence to achieve optimal performance. Many efforts have been made to address this issue by reordering rays based on their origin, end, or direction to enhance coherence. However, a drawback of reordering methods is the need to encode and sort rays before tracing, introducing additional overhead. We propose a technique to generate coherent rays directly by reusing the direction. Additionally, we introduce an interleaved reuse domain partition method to mitigate the impact of sampling correlation resulting from direction reuse. We demonstrate the effectiveness of our approach across various scenes, establishing its superiority over reordering methods.
%-------------------------------------------------------------------------
%  ACM CCS 1998
%  (see https://www.acm.org/publications/computing-classification-system/1998)
% \begin{classification} % according to https://www.acm.org/publications/computing-classification-system/1998
% \CCScat{Computer Graphics}{I.3.3}{Picture/Image Generation}{Line and curve generation}
% \end{classification}
%-------------------------------------------------------------------------
%  ACM CCS 2012
%   (see https://www.acm.org/publications/class-2012)
%The tool at \url{http://dl.acm.org/ccs.cfm} can be used to generate
% CCS codes.
%Example:


\end{abstract}  
%-------------------------------------------------------------------------
\section{Introduction}

Path tracing involves three crucial steps: computing intersections between rays and the scene, shading the ray hits, and generating new rays based on the intersection results. Among these steps, the intersection test between rays and the scene proves to be the most time-consuming segment. To expedite this step, acceleration structures such as bounding volume hierarchies or KD-trees are typically employed.

The intersection tests are computed in parallel by threads on the GPU. Each thread searches the acceleration structure to find the leaf nodes that the ray passes through. However, when threads, sharing the same cache (such as threads in a warp on NVIDIA GPUs), handle rays requiring access to various nodes of the acceleration structure, divergence occurs, leading to cache misses that significantly slow down the tracing speed.

To mitigate divergence and enhance ray coherence, an effective approach is to reorder rays so that the threads sharing the same cache will process rays passing through similar nodes. Ray reordering methods encode and sort rays based on the spatial feature of rays. Existing ray encoding methods encode rays based on their origins and directions, origins and ends, or even the acceleration structure. By grouping rays with similar spatial characteristics, reordering methods enhance ray coherence, thus accelerating ray tracing.

However, reordering methods suffer from two significant drawbacks. Firstly, they necessitate encoding and sorting rays, introducing additional overhead. Secondly, if the rays are sufficiently separate (as in large-scale landscape scenes), regardless of the encoding or sorting method used, they remain incoherent. Consequently, reordering methods do not perform well in landscape scenes.

In response, we propose a direction reusing method to directly generate coherent rays. We gather nearby pixels into groups and categorize these groups into three categories based on the material of the shading points corresponding to the pixels in each group. Different types of groups employ different strategies to reuse the direction of the secondary rays corresponding to the pixels. By reusing directions, the secondary rays corresponding to pixels in the same group exhibit similar directions. Additionally, since the pixels in the same group are nearby, the origins of these rays (i.e. the shading points corresponding to nearby pixels) are also close to each other. Consequently, the secondary rays in the same group become coherent rays.

Furthermore, to mitigate the impact of sampling correlation resulting from direction reusing, we propose an interleaved grouping method to prevent pixels in the same group from being too close together, which could create noticeable artifacts, or too far apart, which could result in their corresponding shading points not being close enough.

On one hand, generating coherent rays directly does not introduce additional overhead and eliminates the need for encoding and sorting rays. On the other hand, it ensures that the secondary rays in the same group possess similar directions.

The experimental results clearly demonstrate that our method outperforms reordering methods in terms of acceleration performance. Furthermore, the artifacts caused by sampling correlation can be eliminated after several iterations.

%-------------------------------------------------------------------------
\section{Relate Works}
%\subsection{Exploit Coherent Rays}
\subsection{Ray Reodering}
Ray reordering methods enhance ray coherence by rearranging rays based on their spatial characteristics. Initially introduced by Pharr and Hanrahan \cite{pharr1997rendering}, this technique involved sorting rays before tracing to optimize cache performance on CPUs. Subsequently, this reordering technique was extended to GPUs. Several reordering methods based on space-filling curves have been proposed. These methods utilize space-filling curves such as the Morton code to encode the spatial features of secondary rays and sort them accordingly. Reis et al. \cite{reis2017coherent} proposed encoding the origin and direction of secondary rays within the code, where the higher bits represent the origin and subsequent bits represent the direction. Costa et al. \cite{costa2015accelerating} utilized the opposite approach, where higher bits represented the direction. Aila and Karras \cite{aila2010architecture} employed bit interleaving to represent both origin and direction. The interleaved bits from the origin and termination point of the ray were also utilized for encoding \cite{moon2010cache,meister2020ray}.

However, reordering methods based on space-filling curves suffer from a drawback known as 'boundary drift,' preventing them from fully benefiting from longer codes. Xiang et al. \cite{xiang2023faster} highlighted this limitation and proposed an encoding method based on the acceleration structure, called the multilevel hierarchy cut code (MLHCC). MLHCC surpasses space-filling curve-based reordering methods; however, it requires access to the acceleration structure, which is often inaccessible in most ray tracing APIs like DXR. Consequently, it is challenging to apply MLHCC to existing path tracers.

Other reordering methods also exist. Boulos et al. \cite{boulos2008adaptive} explicitly rearranged rays into coherent ray packets, while Áfra et al. \cite{afra2016local} sorted rays based on material similarity.

All reordering methods necessitate encoding and sorting rays before tracing, introducing additional overhead. Moreover, reordering methods do not perform effectively in landscape scenes, as the spatial features of rays in such scenes are often disparate.

\subsection{Other Methods to Exploit Coherent Rays}
Reshetov et al. \cite{reshetov2005multi} introduced a method that grouped rays into frusta to minimize the number of intersection tests. Torres et al. \cite{torres2011traversing} further extended this concept by dividing the BVH into a forest of disjoint subtrees based on their cuts. Garanzha and Loop \cite{garanzha2010fast} utilized breadth-first packet traversal to reduce divergence in computations, drawing inspiration from Arvo and Kirk’s 5D ray tracing \cite{arvo1987fast}. However, similar to MLHCC, these methods require accessible acceleration structures, limiting their practicality.

Nimier-David et al. \cite{nimier2019mitsuba} implemented Coherent Pseudo-Marginal MLT (CPMMLT) in Mitsuba 2. CPMMLT generates highly coherent rays, leading to faster intersection computations. It's important to note, however, that CPMMLT comes with the drawback of requiring a specialized rendering pipeline.

\subsection{Modifying Acceleration Structure}

While reordering methods aim to enhance ray coherence and reduce cache misses, some approaches focus on optimizing acceleration structures to achieve similar outcomes. One widely adopted acceleration structure is the Bounding Volume Hierarchy (BVH) \cite{goldsmith1987automatic}, which has been extensively researched and refined \cite{macdonald1990heuristics, lin2020hardware}. As previously mentioned, incoherent rays result in excessive memory access, a concern given the heightened sensitivity of current GPUs to memory traffic. To address this, a solution involving wider trees to minimize memory access has been proposed.

Ylitie et al. \cite{ylitie2017efficient} demonstrated that efficient traversal of incoherent rays can be achieved through a compressed wide BVH, emphasizing the significance of reducing memory consumption within the hierarchy for optimal performance. Similarly, Pharr et al. \cite{pharr1997rendering} divided the scene into cells and computed ray-cell intersections, thereby reducing the necessity for disk-to-memory scene switches, particularly in vastly expansive scenes. Expanding on this concept, Navratil et al. \cite{navratil2007dynamic} extended the approach by suggesting the partitioning of both geometries and rays. Furthermore, Hendrich et al. \cite{hendrich2019ray} introduced a ray classification scheme that maps rays to deeper interior nodes in the tree, effectively bypassing nodes at higher levels. While all these methods modify the acceleration structure to minimize memory access, applying these modifications to path tracers implemented through ray tracing APIs proves challenging.

%-------------------------------------------------------------------------
\section{Method}
\subsection{Directly Direction Reusing} 

Ray reordering methods have demonstrated that sorting rays based on their origins and directions enhances ray coherence, resulting in accelerated ray tracing. Consequently, generating coherent rays can be simplified to producing rays with similar origins and directions.

To elucidate our method, we assume that each pixel samples a single path (the scenario where each pixel samples multiple paths can be considered as sampling one path in smaller pixels). Thus, each pixel corresponds to a shading point, which represents the first intersection of the path with the scene. The direction of the primary ray on the path is the incidence direction at the shading point, while the direction of the secondary ray represents the exit direction.

When considering the secondary rays, if the pixels corresponding to these rays are in close proximity, it is likely that the rays' origins are also nearby. Our goal is to ensure these rays have similar directions. A straightforward approach is to make these rays share the same direction.

The 'Direct Direction Reusing' method divides the screen into several square tiles and randomly selects one pixel along with its corresponding secondary rays within each tile. Only the selected ray samples a direction, while the other rays in the same tile reuse this direction in world space, calculating the probability based on the direction and their respective materials. Consequently, the secondary rays within each tile share the same directions

\begin{figure}[htb]
\centering
% the following command controls the width of the embedded PS file
% (relative to the width of the current column)
\includegraphics[width=.8\linewidth]{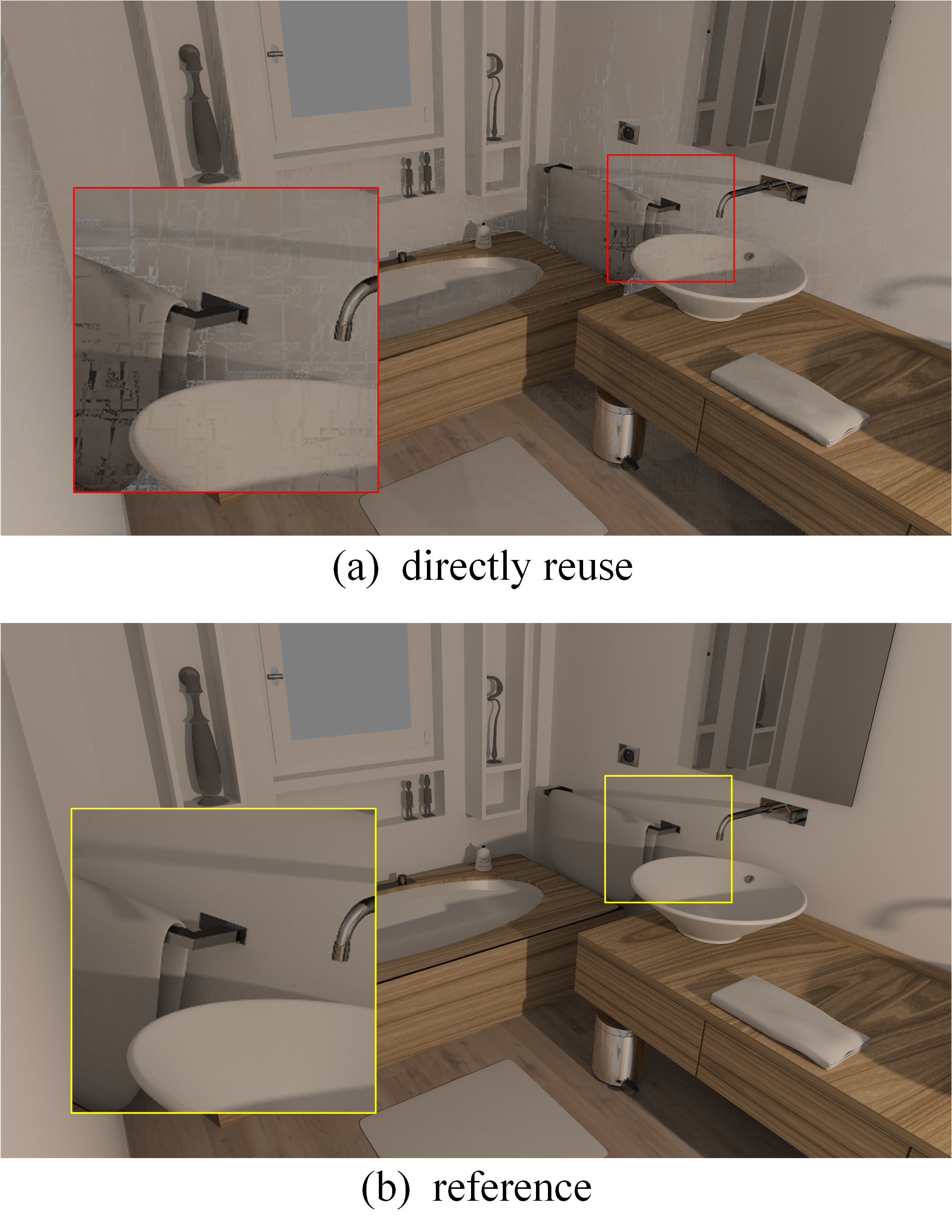}
% replacing the above command with the one below will explicitly set
% the bounding box of the PS figure to the rectangle (xl,yl),(xh,yh).
% It will also prevent LaTeX from reading the PS file to determine
% the bounding box (i.e., it will speed up the compilation process)
% \includegraphics[width=.95\linewidth, bb=39 696 126 756]{sampleFig}
%
\parbox[t]{.9\columnwidth}{\relax

}
\caption{\label{fig:directReuse}
Directly Direction Reuse. Comparing with reference, the artifacts usually occur in the tiles that contain shading points which have different normal. The exit direction, which is shared by these shading points, even points to the back face for some of the shading points. }
\label{directReuse}
\end{figure}

Directly reusing directions doubles the speed of secondary ray tracing, showcasing that generating coherent rays directly can indeed expedite ray tracing. However, this approach leads to noticeable artifacts in the rendering results (Fig. \ref{directReuse}). These artifacts arise because direct direction reusing violates the principle of importance sampling.

On one hand, the hemispherical surfaces corresponding to the shading points located on two nonparallel planes are not identical. Reusing the sampling direction of one point for another point will result in the latter being unable to sample its complete corresponding hemispherical surface.

On the other hand, unless spherical uniform sampling is employed, points situated on two nonparallel planes with distinct materials do not possess the same probability of sampling the same direction. When the probability of sampling the corresponding direction is computed at a point reusing the direction sampled from another point, it leads to an incorrect calculation of the sampling probability.

%---------------------------------------------
\subsection{Tangent Space Direction Reusing}\label{section:ts}
\begin{figure*}[tbp]
\centering
\includegraphics[width=.8\linewidth]{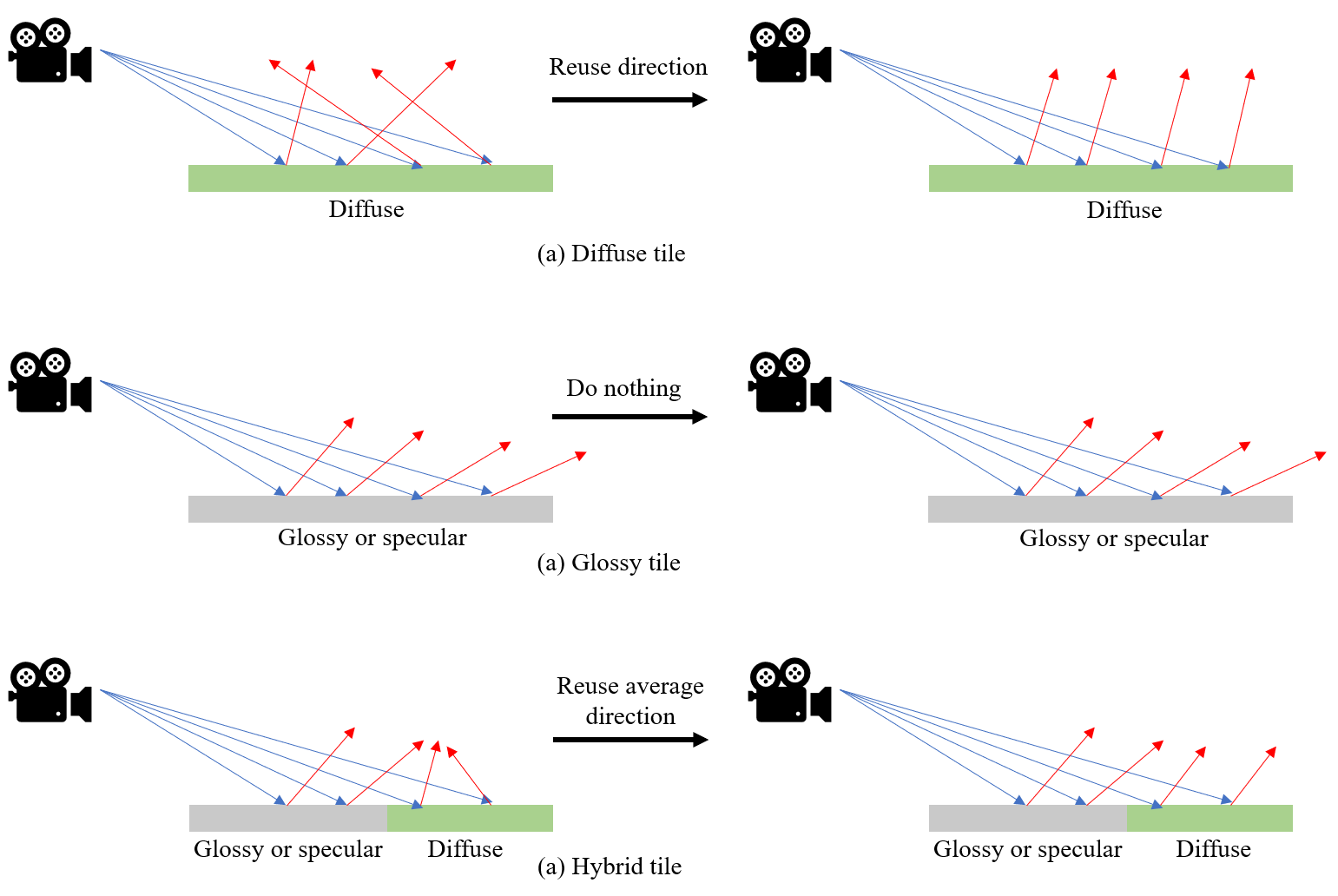}

\parbox[t]{.9\columnwidth}{\relax

}
\caption{\label{fig:tileType}
Strategy of Different Tile Types. (a) Reuse the exit direction of one of the shading points in diffuse tile. (b) Do nothing in glossy tile. (c) Reuse the average exit direction of glossy or specular shading points in hybrid tile.}
\label{tileType}
\end{figure*}

To mitigate artifacts, we opt to reuse directions in tangent space rather than world space. By reusing directions in tangent space, secondary rays belonging to the same tile share a common direction, provided their origins are on parallel or similar planes. If their origins lie on nonparallel planes, the direction will be adaptively adjusted based on the surface normal. This approach ensures that we avoid sampling the lower hemisphere and accurately sample the complete sample space at each shading point.

Furthermore, we classify tiles into three types based on the material of the shading points corresponding to the contained pixels. Each type is handled with a different strategy as figure.\ref{tileType} shown.

\textbf{Diffuse Tile}: This type consists of shading points with only diffuse materials. Since the sampling probability of diffuse material is direction-independent, reusing the tangent space direction does not violate the principle of importance sampling.

\textbf{Glossy Tile}: Tiles with only glossy or specular shading points undergo no processing. The primary objective of direction reuse is to ensure that the exit directions of shading points (i.e., the direction of secondary rays) within the same tile are similar. Therefore, since shading points within the same tile typically have similar positions and incident directions when the material is glossy or specular, there is no necessity to modify their directions.

\textbf{Hybrid Tile}: These tiles contain a mix of both diffuse and glossy or specular shading points. We sample the exit directions for all the glossy or specular shading points, calculate their average direction in tangent space, and then set the exit directions of the diffuse shading points to this average direction. As analyzed earlier, the exit directions of glossy shading points are coherent, and only the exit directions of diffuse shading points need modification. Thus, setting them to the average exit direction of glossy shading points generates a beam of coherent rays.

%-------------------------------
\subsection{Interleaved Grouping}

Direction reusing involves sharing the sampling direction among pixels in a tile, leading to sampling correlation. While it's impossible to completely eliminate sampling correlation due to the necessity of direction reusing, we propose an 'Interleaved Grouping' approach to mitigate its impact.

\begin{figure}[htb]
\centering

\includegraphics[width=.9\linewidth]{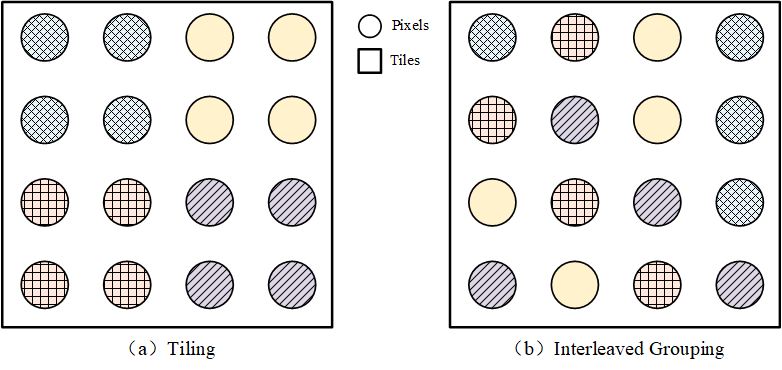}

\parbox[t]{.9\columnwidth}{\relax

}
\caption{\label{fig:interleavedGrouping}
Interleaved Grouping. The pixles with same color in the figure share the secondary rays' direction.}
\label{interleavedGrouping}
\end{figure}

The Interleaved Grouping method divides the pixels on the screen into several W*H tiles. Within each tile, pixels are numbered based on their order on the screen. These numbers are then shuffled using a linear congruence method (only the pixel numbers are shuffled, not the pixels themselves). Subsequently, the pixels in each tile are divided into WH/K groups, as depicted in Fig. \ref{interleavedGrouping}. If the shuffled numbers of pixels, divided by K, yield the same quotient, the pixels are grouped together. Pixels within the same group share the secondary ray's direction using the method outlined in Section \ref{section:ts}.

This approach offers a balance: reusing direction within an interleaved group instead of within a tile ensures that pixels sharing the same secondary ray's direction are not too close, preventing noticeable artifacts, nor too far to share similar incident direction. Furthermore, due to the random shuffling, the composition of interleaved groups varies in each tile and in each frame. Consequently, after several iterations, the artifacts gradually disappear.

%-------------------------------------
\begin{table*}[]
\centering
\caption{Secondary ray trace speed comparison of our methods and test methods. The trace speed is measured in million rays per second (MRay/s). The numbers in parentheses represent the acceleration ratio of the method.}
\label{tab_Results_Overview}
%\resizebox{\linewidth}{26mm}{
\begin{tabular}{lcccccc}
\hline
Scene         & Breakfast  & Salle de Bain & Living room  & Sibenik     & Vokselia   \\ \hline % & MultiInstance & San   \\ \hline
\#Triangles   &     1347K       &       1231K        &      580K       &     75K        &     1876K   \\ \hline % & 309K & 5.6M   \\ \hline
&    
\begin{minipage}[b]{0.3\columnwidth}
\centering
\raisebox{-.5\height}{\includegraphics[width=\linewidth]{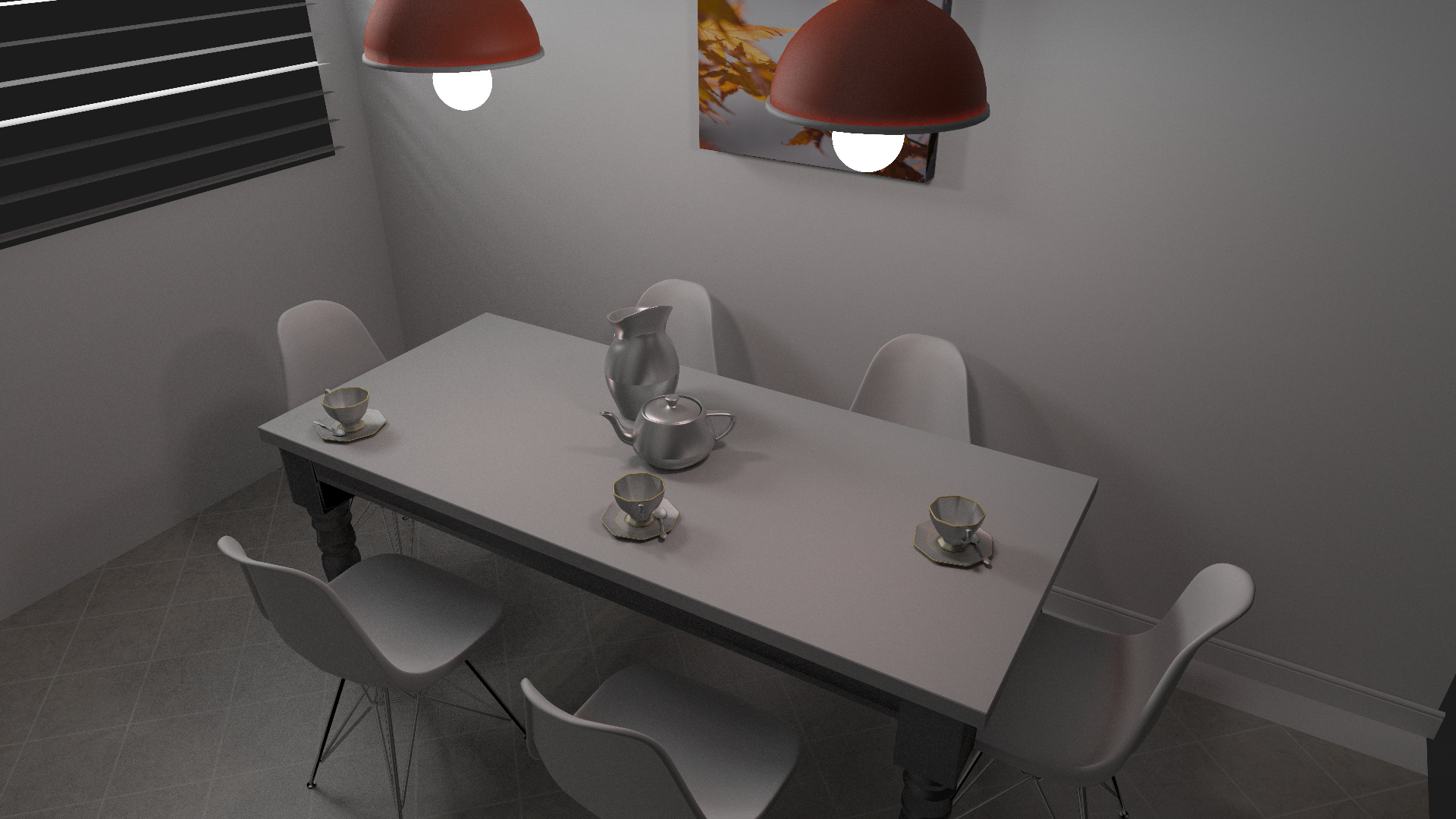}}
\end{minipage}         &    
\begin{minipage}[b]{0.3\columnwidth}
\centering
\raisebox{-.5\height}{\includegraphics[width=\linewidth]{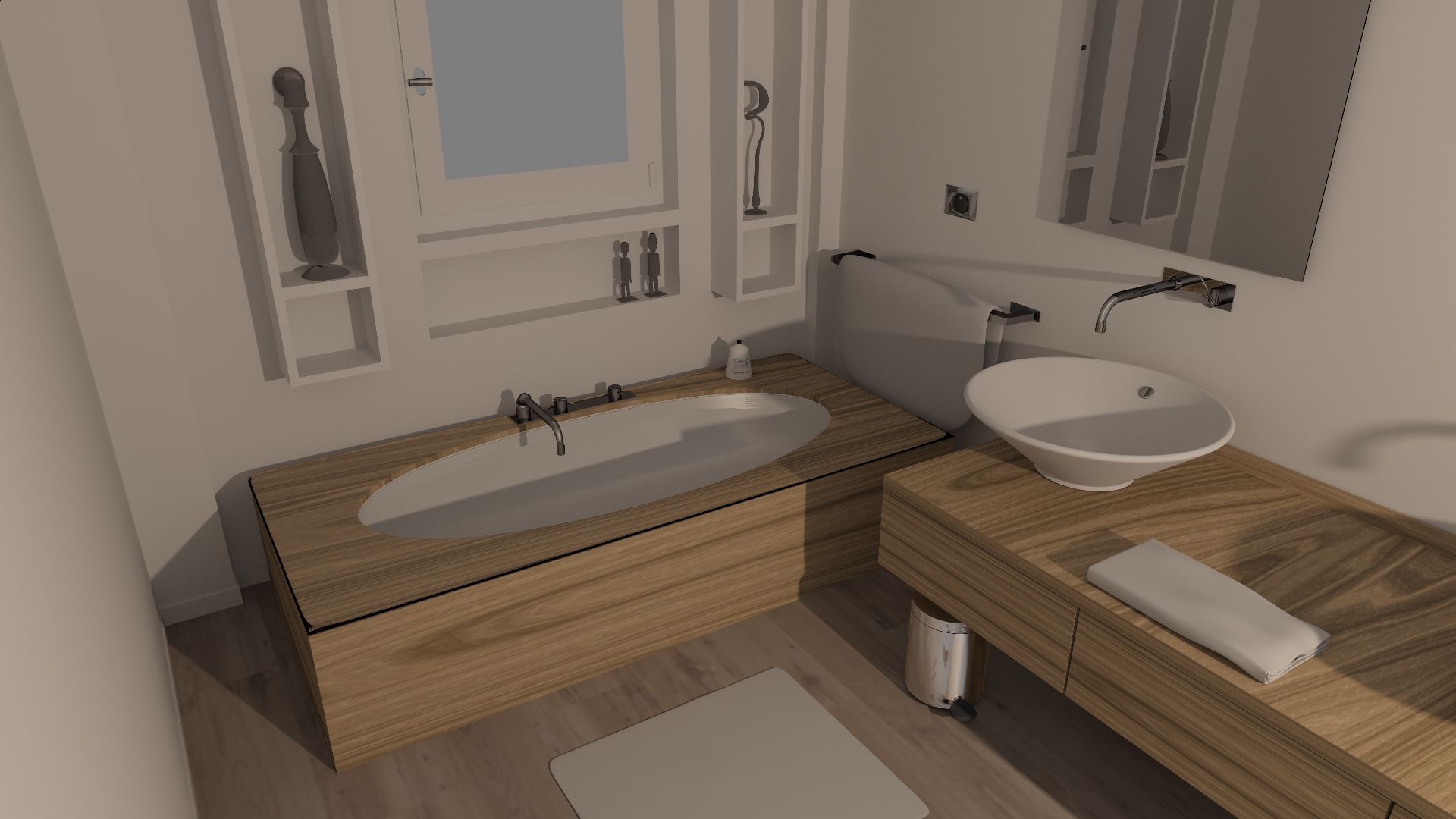}}
\end{minipage}
&         
\begin{minipage}[b]{0.3\columnwidth}
\centering
\raisebox{-.5\height}{\includegraphics[width=\linewidth]{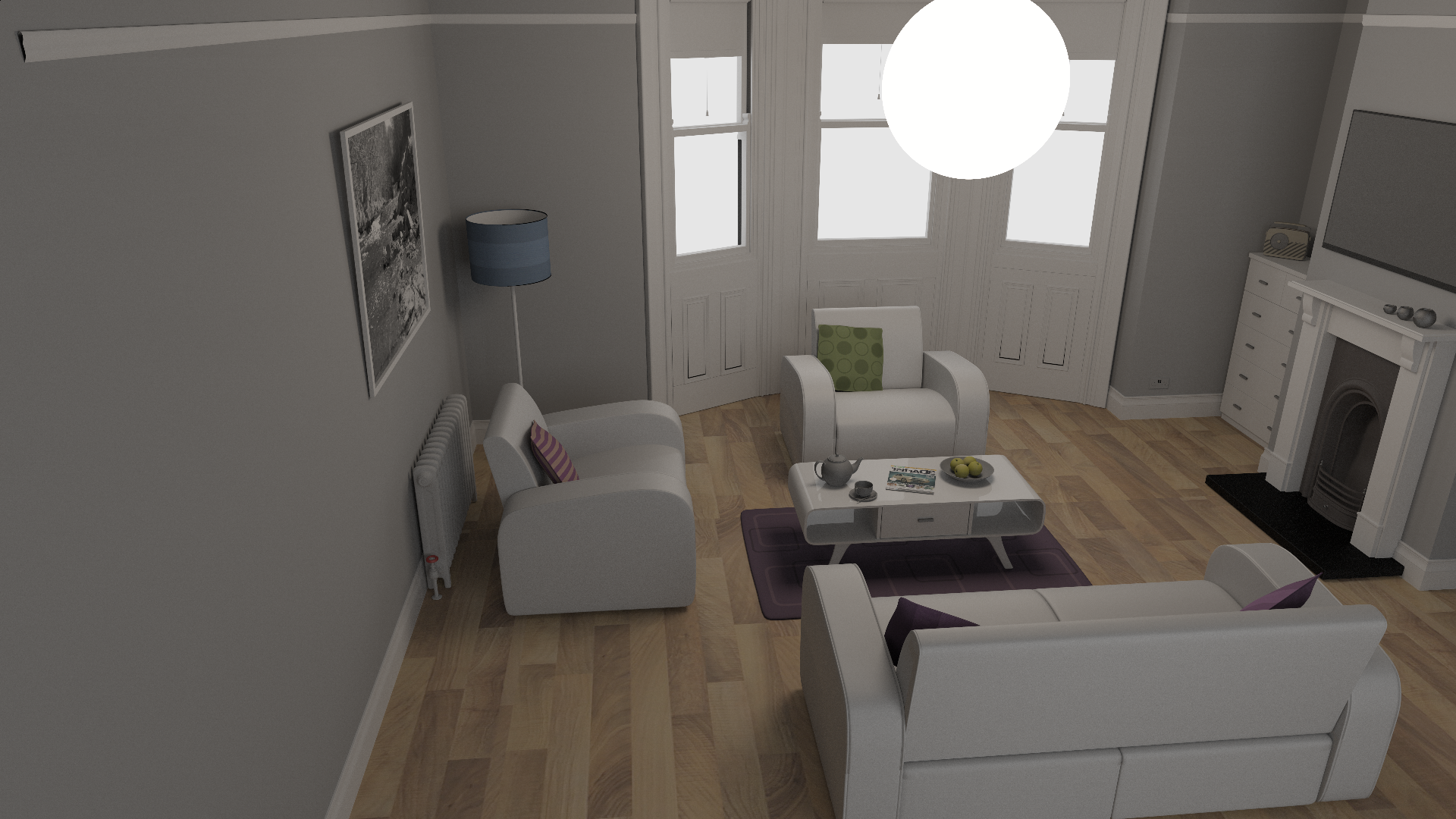}}
\end{minipage}
&      
\begin{minipage}[b]{0.3\columnwidth}
\centering
\raisebox{-.5\height}{\includegraphics[width=\linewidth]{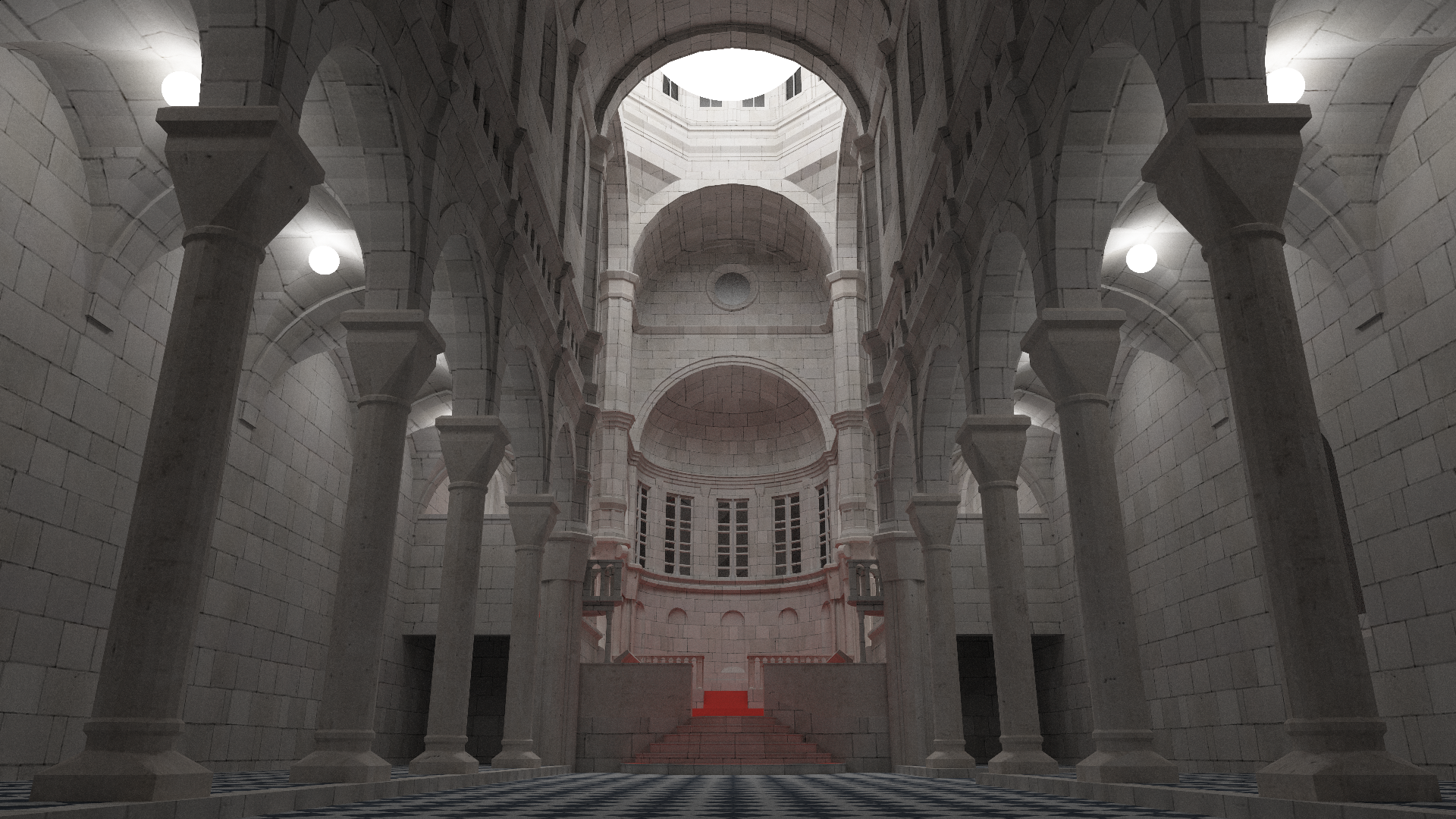}}
\end{minipage}
&    
\begin{minipage}[b]{0.3\columnwidth}
\centering
\raisebox{-.5\height}{\includegraphics[width=\linewidth]{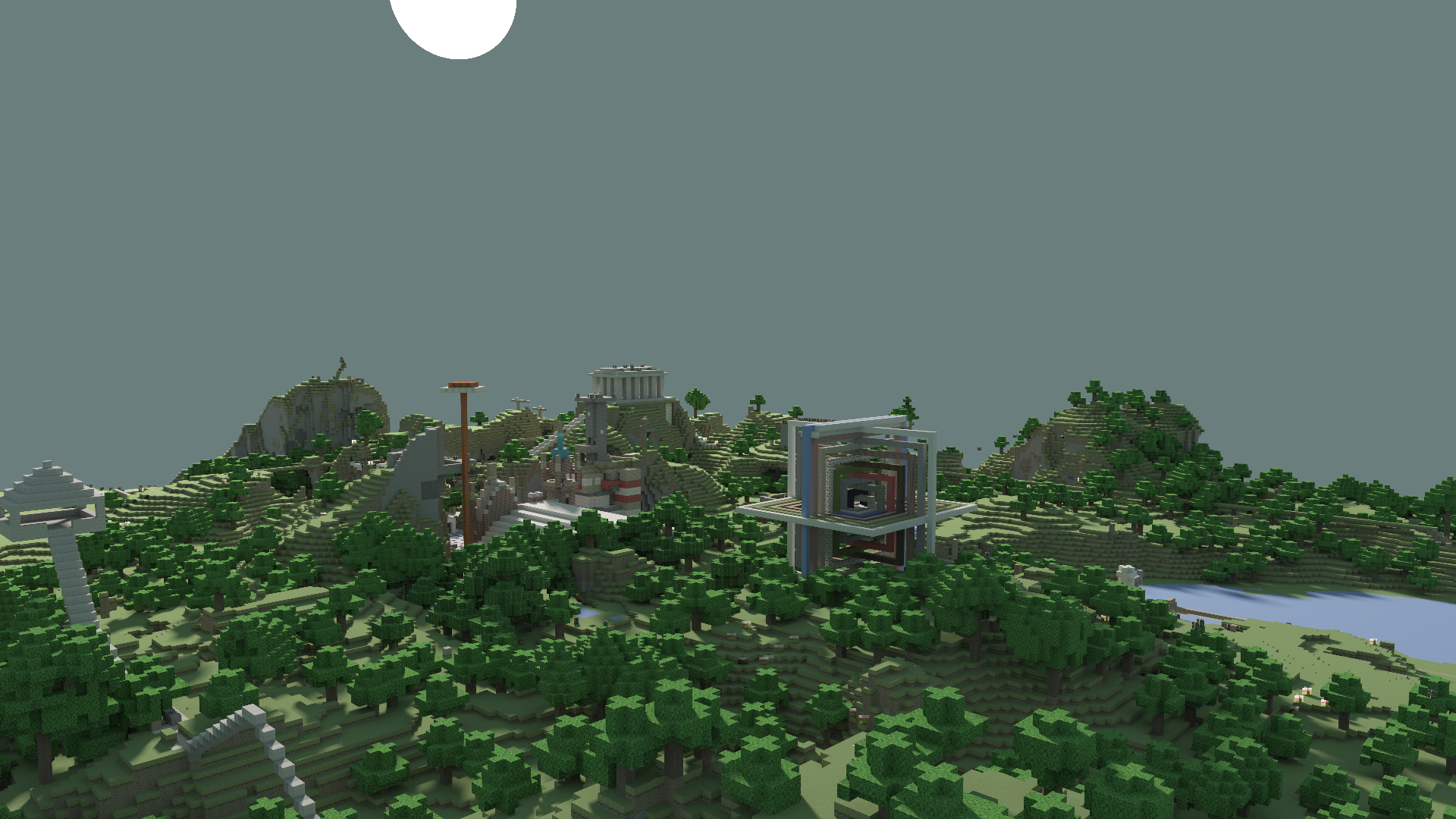}}
\end{minipage}
%&    
%\begin{minipage}[b]{0.2\columnwidth}
%\centering
%\raisebox{-.5\height}{\includegraphics[width=\linewidth]{img/MultiInstance}}
%\end{minipage}
%&    
%\begin{minipage}[b]{0.2\columnwidth}
%\centering
%\raisebox{-.5\height}{\includegraphics[width=\linewidth]{img/San}}
%\end{minipage}
\\ \hline
NoOpt      & 37.4 (1.00) & 206.1 (1.00)   & 189.4 (1.00) & 154.7 (1.00) & 64.2 (1.00)\\ % & 209.5(1.00)& 32.7(1.00)\\
Ours    & 58.6(1.57) & \textbf{305.6(1.47)}   & \textbf{287.9(1.50)} & \textbf{278.5(1.80)} & \textbf{87.5(1.36)}\\ % & \textbf{264.2(1.26)} & \textbf{40.7(1.24)}\\
MLHCC32		& \textbf{65.1 (1.74)} & 280.2 (1.36)   & 277.8 (1.47) & 218.5 (1.41) & 74.4 (1.16)\\ %& 275.0(1.31)&39.8(1.21) \\
TwoPoint    & 56.8 (1.52) & 274.2 (1.33)   & 272.1 (1.44) & 217.9 (1.41) & 75.5 (1.18)\\ %& 242.0(1.15)& 39.5(1.21)\\
AilaCompact & 55.6 (1.48) & 268.0 (1.30)   & 258.2 (1.36) & 208.9 (1.35) & 75.5 (1.18)\\ %& 195.7(0.93)& 38.7(1.18) \\
Reis          & 49.7 (1.33) & 242.4 (1.18)   & 224.8 (1.19) & 183.2 (1.18) & 74.3 (1.16)\\ %& 202.2(0.97)& 37.0(1.13) \\
Origin        & 46.1 (1.23) & 233.0 (1.13)   & 208.3 (1.10) & 165.4 (1.07) & 74.8 (1.17)\\ \hline %& 199.0(0.95)& 35.1(1.08)\\ \hline

\end{tabular}
\end{table*}

\begin{figure*}[tbp]
\centering
\includegraphics[width=\linewidth]{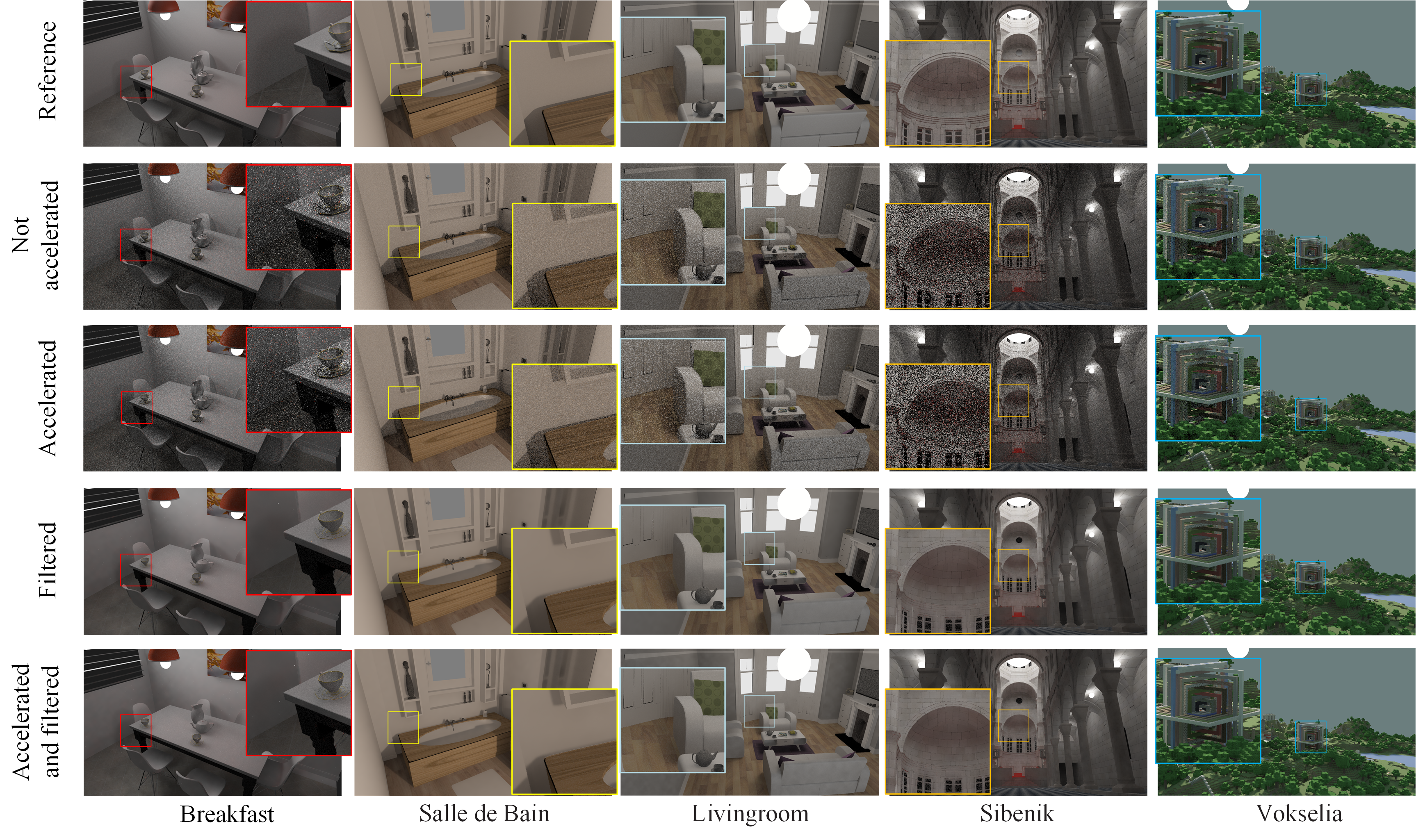}

\parbox[t]{\columnwidth}{\relax

}
\caption{\label{fig:renderingResult}
Rendering Result.}
\label{renderingResult}
\end{figure*}

\section{Implementation}
%TODO
To compare the performance of our technology with existing reordering methods, we select five representative methods: MLHCC\cite{xiang2023faster} and Origin, Reis, AilaCompact, TwoPoint as introduced by Meister et al.\cite{meister2020ray}. Among the various termination point estimation methods in the TwoPoint method, we chose the fixed-length estimation method, which sets the ray length to 0.25 of the largest extent of the scene bounding box. The code lenth of MLHCC is 32. We implemente our method in CUDA C on NVIDIA GeForce RTX 3060 GPU using the same framework as MLHCC. We also implemente our method by DXR on NVIDIA GeForce RTX 3070 GPU using Falcor\cite{Kallweit22}.

\section{Result}
The performance is evaluated in five scenes of various complexity. Thanks to McGuire \cite{McGuire2017Data} and Bitterli \cite{resources16} for providing the necessary models. The measurements were performed on a GeForce RTX 3060 GPU with an image resolution of 1920 x 1080. The code was compiled using CUDA 10.0. All recorded data represents averages. Each scene was run three times, and the data from 100 frames were collected to calculate the average running time and the number of rays. We set $\lambda$ = 0.125 for MLHCC, average distance = 0.25 for TwoPoint method. For our method, we set tile size to 8x8 and group size to 8.

\subsection{Acceleraton Performance}
As illustrated in Table \ref{tab_Results_Overview}, our method has consistently demonstrated improved ray tracing efficiency across various scenes, including densely populated indoor settings (Bathroom, Breakfast), expansive indoor environments (Sibenik), and extensive outdoor landscapes (Vokselia). Overall, our method achieved a speedup ranging from 1.4 to 1.8 times, with the specific speedup ratio influenced by occlusion within the scene.

Notably, in scenes with significant occlusion, where a nearby object obstructs a distant object, the assumption that shading points for pixels within the same tile are close to each other becomes unreliable. Consequently, this affects the acceleration effect, which explains why our method did not achieve optimal results in the Breakfast scene. Similarly, while our method outperforms other reordering methods in Salle de Bain and Living Room, the acceleration ratios do not show as significant an improvement as observed in Sibenik and Vokselia, owing to the challenge posed by large-scale occlusion.

Furthermore, our method excels in landscape scenes like Vokselia compared to reordering methods. Reordering methods encode rays before sorting, but landscape scenes are rich in details and cannot be accurately represented by a 32-bit code. Even with ray reordering based on the code, ensuring ray coherence is difficult. In contrast, our method directly generates coherent rays by reusing directions, resulting in rays within the same group being inherently coherent, or at least possessing similar directions.

Moreover, our method only shares directions between threads using shared memory, making the extra overhead negligible. On the contrary, reordering methods necessitate ray encoding and sorting, constituting approximately 5\% of the overall processing time \cite{xiang2023faster}.

%--------------------------------------
\subsection{Image Quality}
Figure \ref{renderingResult} showcases the rendering outcomes for each scene. The top row exhibits the reference images, each generated after 4096 iterations. The third and fifth rows demonstrate the application of our method, while the fourth and fifth rows depict the utilization of the SVGF\cite{schied2017spatiotemporal} filtering strategy. For every scene, a 256x256 pixel area was designated as a hotspot region, outlined with a colored border, and subsequently magnified to facilitate comparison. A comparison between the second and third rows, as well as the fourth and fifth rows, allows for a clear observation of the alterations in rendering effects pre and post our method's implementation. It's important to note that, with appropriately configured parameters, our method does not markedly impact the rendering results.

\begin{table}[]
\centering
\caption{The PSNR of the images before and after applying our method after 128 iterations.}
\label{table43}
\resizebox{\linewidth}{5mm}{
\begin{tabular}{lcccccccc}
\hline

Scene         & Breakfast  & Salle de Bain & Living room  & Sibenik     & Vokselia    \\ \hline
PSNR(Ours)  &20.91	&24.61	&19.72	&19.61	&33.72 \\
PSNR(path tracing) &20.94	&24.65	&19.81	&19.63	&33.79 \\
\hline
\end{tabular}}
\end{table}

To quantitatively evaluate the quality of the rendered images, we employ the Peak Signal-to-Noise Ratio (PSNR). Table \ref{table43} displays the PSNR of the images before and after the application of our method after 128 iterations. It's evident that there is no substantial difference in PSNR between the two sets of images across all the test scenes.

%--------------------------------------
\subsection{Tile Size and Group Size}
The group size K and tile size (W, H) significantly impact the efficiency and rendering quality of our method. Table \ref{table44} illustrates the influence of different group sizes and tile sizes on the ray tracing speed in the Sibenik scene, with data representing the speed of tracing all secondary rays. Notably, a smaller tile size (W, H) leads to faster ray tracing, and a larger group K results in faster ray tracing as well. Employing smaller tiles and larger groups achieves a higher acceleration, aligning with the fundamental principle of generating coherent rays. Larger groups imply that more rays share similar directions, whereas smaller tiles ensure that the origins of rays are closer to each other.

\begin{figure*}[tbp]
\centering
\includegraphics[width=.9\linewidth]{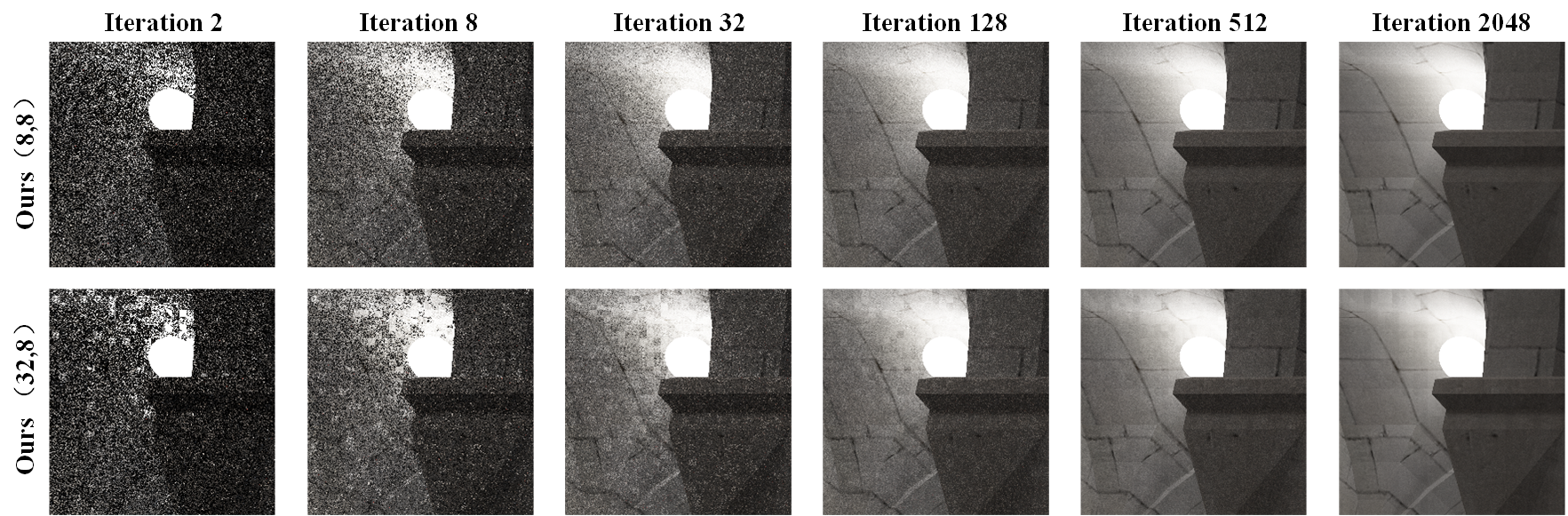}

\parbox[t]{\columnwidth}{\relax

}
\caption{\label{fig:GuideComp_8_32}
The rendering effects for K=8 and K=32, both rendered with a tile size of (8, 8). When K approaches the number of pixels within a tile (W*H), noise becomes prominent with fewer iterations.}
\label{GuideComp_8_32}
\end{figure*}

However, it's essential to emphasize that the maximum group size K and the minimum tile size (W, H) may not always be the best parameter choices. Excessively small tiles and overly large groups can both adversely affect rendering quality. Figure \ref{GuideComp_8_32} offers a comparison of the rendering effects for K=8 and K=32, both rendered with a tile size of (8, 8). It's evident that as K approaches the number of pixels within a tile (W*H), noise becomes prominent with fewer iterations. Even after accumulating up to 2048 iterations, such noise persists and accumulates into noticeable pseudoglyphs.

\begin{figure*}[tbp]
\centering
\includegraphics[width=\linewidth]{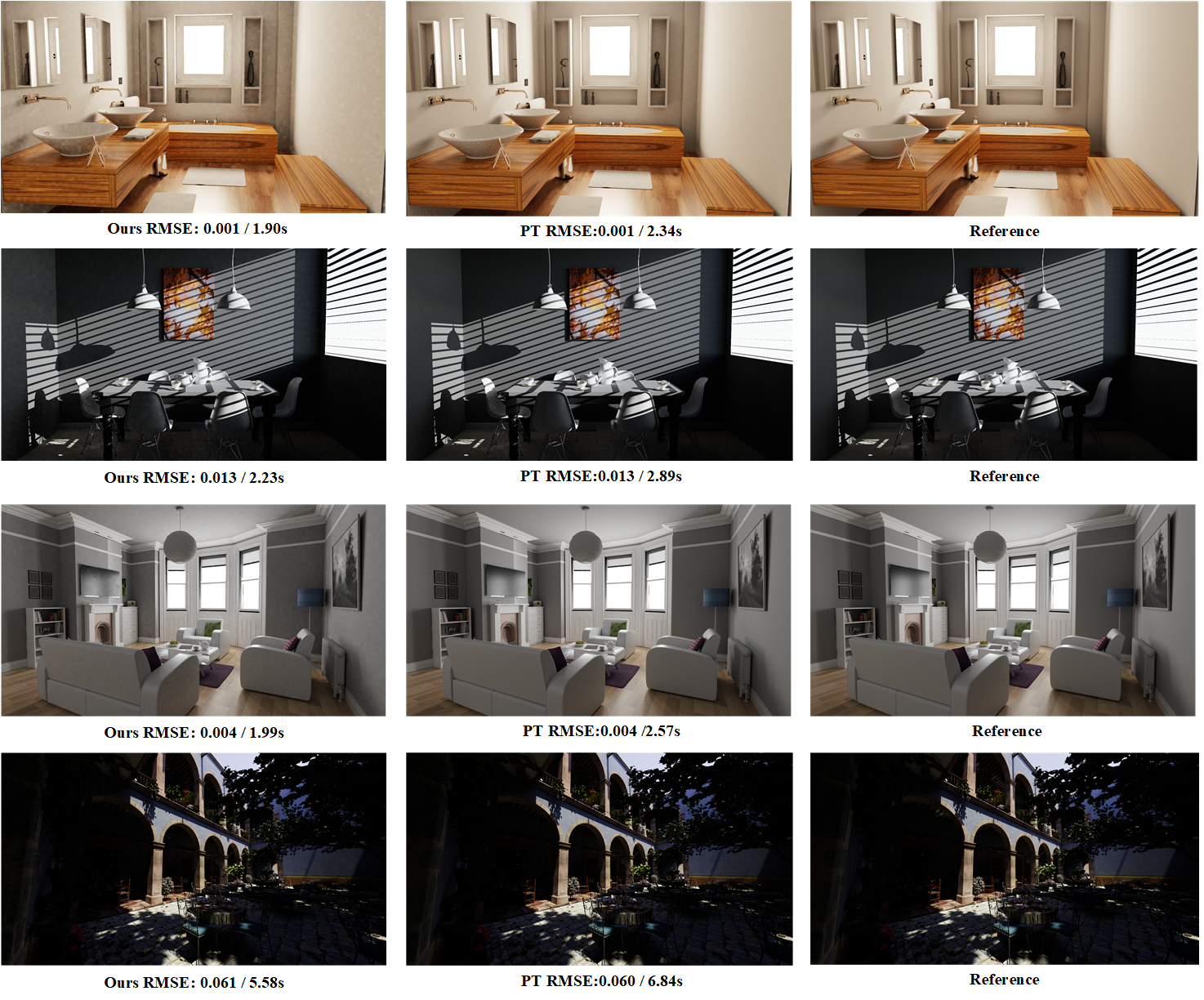}

\parbox[t]{\columnwidth}{\relax

}
\caption{\label{fig:GuideComp_O_8}
The rendering result of the implemention on DXR. To generate the images which have nearly the same RMSE, the path tracer using our method cost less time than traditional one.}
\label{GuideComp_O_8}
\end{figure*}

\begin{table}[]
\centering
\caption{The impact of different group sizes and tile sizes on ray tracing speed in Sibenik scene. The data represent trace speed of all secondary rays.}
\label{table44}
\resizebox{\linewidth}{10mm}{
\begin{tabular}{lcccccccc}
\hline
(W,H)     & K = 8  & K = 16 & K = 32  &  K = 64 \\ \hline
(8,8) &278.5 (1.80)	&303.9 (1.97)	&378.0 (2.45)	&379.4 (2.46)\\
(16,8) &234.1 (1.52)	&290.4 (1.88)	&338.9 (2.19)	&337.7 (2.19)\\
(16,16) &232.5 (1.51)	&273.9 (1.77)	&305.2 (1.98)	&306.6 (1.99)\\
(32,8) &227.1 (1.47)	&255.0 (1.65)	&298.9 (1.94)	&298.5 (1.93)\\

\hline
\end{tabular}}
\end{table}

\subsection{Application on DXR}
Our method seamlessly integrates with any GPU-based path tracing framework. We made modifications to Falcor's path tracer, which is built on DXR, to implement our approach. The testing was performed on an NVIDIA GeForce RTX 3070 GPU, with an image resolution set to 1920 x 1080. We opted for a tile size of 16x16 and a group size of 32. As illustrated in Figure \ref{GuideComp_O_8}, our method notably reduces processing time while maintaining comparable RMSE to traditional path tracing.

%------------------------------------------------------------
\section{Conclusion and Future Work}

We present a directional reusing method to directly generate coherent rays. In comparison to reordering methods that require encoding and sorting rays for coherency, our approach offers three significant advantages. Firstly, our method only requires reusing the directions of secondary rays, minimizing additional overhead. Secondly, it demonstrates superior acceleration performance compared to reordering techniques. Thirdly, our approach showcases greater robustness, particularly in extensive landscape scenes. While it's true that the sampling correlation resulting from direction reusing can sometimes introduce artifacts when our method is applied, experiments have shown that these artifacts tend to diminish after a few iterations.

It's important to note that our method currently does not handle transmission materials. When rays scatter through transmission materials, the exit directions point towards the back end, necessitating sampling of the lower hemisphere. This behavior significantly differs from that of diffuse and glossy materials. Extending our method to accommodate transmission materials would be a valuable endeavor, potentially leading to enhanced performance.

%-------------------------------------------------------------------------
% bibtex
\bibliographystyle{eg-alpha-doi} 
\bibliography{egbibsample}       

% biblatex with biber
% \printbibliography                

%-------------------------------------------------------------------------

%---------------------------------------------------------------

\end{document}